\documentclass[twocolumn,showpacs,aps,dvipdmx]{revtex4}
\usepackage{graphicx}
\usepackage{bm}
\usepackage{amsmath}
\usepackage{amssymb}
\usepackage{color}

\def\l{\langle}
\def\r{\rangle}

\begin{document}
\title{
Entropy of the diluted antiferromagnetic Ising models 
on the frustrated lattices \\
using the Wang-Landau method
} 
\author{Yuriy Shevchenko$^{1,2}$}
\email{shevchenko.ya@dvfu.ru}
\author{Konstantin Nefedev$^{1,2}$}
\email{nefedev.kv@dvfu.ru}
\author{Yutaka Okabe$^{3}$}
\email{okabe@phys.se.tmu.ac.jp}
\affiliation{
$^1$School of Natural Sciences, Far Eastern Federal University, Vladivostok, 
Russian Federation \\
$^2$Institute of Applied Mathematics, Far Eastern Branch, 
Russian Academy of Science, Vladivostok, Russian Federation \\
$^3$Department of Physics, Tokyo Metropolitan University, Hachioji, 
Tokyo 192-0397, Japan \\
}

\date{\today}

\begin{abstract}
We use a Monte Carlo simulation to study the diluted antiferromagnetic 
Ising model on the frustrated lattices including the pyrochlore lattice 
to show the dilution effects. 
Using the Wang-Landau algorithm, 
which directly calculates the energy density of states, 
we accurately calculate the entropy of the system. 
We discuss the nonmonotonic dilution concentration dependence 
of residual entropy for the antiferromagnetic Ising model 
on the pyrochlore lattice, 
and compare it to the generalized 
Pauling approximation proposed by Ke {\it et al.} 
[Phys. Rev. Lett. {\bf 99}, 137203 (2007)].
We also investigate other frustrated systems, 
the antiferromagnetic Ising model on the triangular lattice 
and the kagome lattice, demonstrating the difference in 
the dilution effects between the system on the pyrochlore lattice 
and that on other frustrated lattices.

\vspace{1em}
\noindent{This paper is accepted for publication in Phys. Rev. E}
\end{abstract}

\pacs{
05.50.+q, 75.40.Mg, 75.50.Lk, 64.60.De
}

\maketitle

\section{Introduction}

Frustration plays an important role in several scientific fields. 
The simplest example is the antiferromagnetic (AFM) Ising model 
on the triangular lattice.  One cannot select 
all three pairs of spins antiparallel in the basic unit 
of the triangle. It is called frustration. Because of frustration, 
there is no long-range order, and the frustration 
leads to a high degeneracy of ground states. 
The existence of the residual entropy was discussed 
by Pauling in 1935 for water ice \cite{Pauling}. 

Recently, spin-ice materials have captured particular attention 
\cite{Harris,Ramirez,Bramwell}, and their exotic physics 
is a current topic of geometrically frustrated magnets. 
Prototype materials are the pyrochlores Dy${_2}$Ti${_2}$O${_7}$ and 
Ho${_2}$Ti${_2}$O${_7}$. 
In these materials, the magnetic ions (Dy$^{3+}$ or Ho$^{3+}$) occupy 
a pyrochlore lattice of corner-sharing tetrahedra, and the 
local crystal field environment causes the magnetic moments 
to orient along the directions connecting the centers of two 
tetrahedra at low temperatures \cite{Bramwell,Diep}. 
In the low-temperature spin-ice state, the magnetic
moments are highly constrained locally and obey the so-called
``ice rules": two spins point in and two spins point out of each
tetrahedron of the pyrochlore lattice.
This two-in-two-out spin configuration resembles 
the situation of hydrogen atoms in water ice.
The measured residual entropy is very close to Pauling's estimate,
$(1/2)\ln(3/2)$ R, where R is the molar gas constant \cite{Ramirez}.

The dilution effects on frustration were studied by Ke {\it et al.} 
\cite{Ke} for spin-ice materials.  The magnetic ions 
Dy$^{3+}$ or Ho$^{3+}$ are replaced by nonmagnetic Y$^{3+}$ ions. 
Nonmonotonic zero-point entropy as a function of dilution concentration 
was observed experimentally, and a generalization of Pauling's 
theory was discussed \cite{Ke}. More recently, the detailed 
experimental studies combined with Monte Carlo simulations 
were reported \cite{Lin,Scharffe}.

The effect of a magnetic field is another topic of spin-ice 
materials, and recently, Peretyatko, Nefedev, and Okabe \cite{Peretyatko}
studied the effect of a magnetic field on diluted spin-ice materials 
in order to elucidate the interplay of dilution and magnetic field. 
They observed five plateaus in the magnetization curve of the diluted 
nearest-neighbor spin-ice model on the pyrochlore lattice 
when a magnetic field was applied in the [111] direction. 
This effect contrasts with the case of a pure (i.e.~undiluted) model, 
which displays two plateaus. 

In this paper, motivated by the current interest in the pyrochlore 
lattice, we study the entropy of the diluted AFM Ising model 
on the frustrated lattices using the Monte Carlo simulation. 
If one uses a canonical Monte Carlo simulation such as 
the Metropolis algorithm, the estimate of the entropy is made 
by the numerical integration of specific heat.  
A more straightforward way of computing entropy is 
to use the Monte Carlo method that directly calculates 
the energy density of states (DOS) $g(E)$, 
for example, the Wang-Landau (WL) method \cite{WL}. 
The WL method is an efficient algorithm to
calculate $g(E)$ with high accuracy.
Several recent progresses have been made in connection with the WL method.  
To improve the convergence, the $1/t$ algorithm was 
proposed~\cite{Belardinelli}. 
Using the WL method, the difference of the energy DOS was examined 
to discuss the behavior of the first-order transition~\cite{komura12}.
A parallel WL method based on the replica-exchange 
framework for Monte Carlo simulations was also proposed~\cite{vogel13}. 
Recently, Ferreyra {\it et al.} \cite{Ferreyra} reported the calculation 
of $g(E)$ for the Ising model on the pyrochlore lattice 
using the WL method.

As a theoretical model of the spin-ice material, we treat 
the nearest-neighbor AFM Ising model on the pyrochlore lattice. 
A more complicated model, such as the dipolar model, 
may be required to make connections to actual materials.
However, Isakov {\it et al.} \cite{Isakov} discussed 
the reason why the low-temperature entropy of the spin-ice 
compounds is well described by the nearest-neighbor AFM 
Ising model on the pyrochlore lattice, i.e., by the "ice rules". 

The pyrochlore lattice can be regarded as alternating kagome 
and triangular layers, and the magnetic field in the [111] 
direction effectively decouples these layers. 
The triangular and kagome lattices provide the two-dimensional 
frustrated systems. It is interesting to compare 
the behavior of the dilution effects on the AFM Ising model 
of the pyrochlore lattice and those of the triangular 
and kagome lattices.

The purpose of the paper is as follows.  
First, we obtain precise estimates of entropy for diluted systems 
using the WL method. 
Second, we compare the dilution concentration dependence of 
the residual entropy of the pyrochlore lattice 
with the generalized Pauling's estimate by Ref.~\cite{Ke}.
Third, we uncover the mechanism of the nonmonotonic behavior 
of residual entropy. 
Finally, we compare the dilution effects on the residual entropy 
for several frustrated Ising systems on the pyrochlore, 
triangular and kagome lattices.

\begin{figure}
\begin{center}
\includegraphics[height=6.3cm]{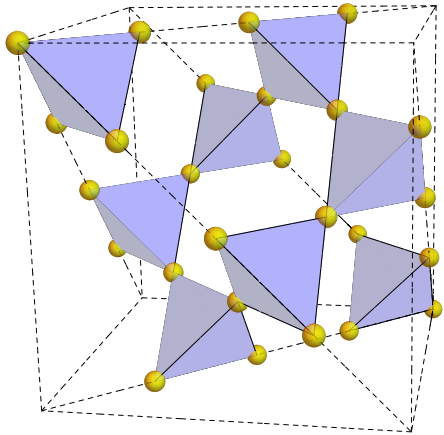}
\caption{
(Color online) 
The illustration of the 16-site cubic unit cell 
of the pyrochlore lattice ($L=1$).
}
\label{fig:fig1}
\end{center}
\end{figure}

\begin{figure}
\begin{center}
\includegraphics[width=8.4cm]{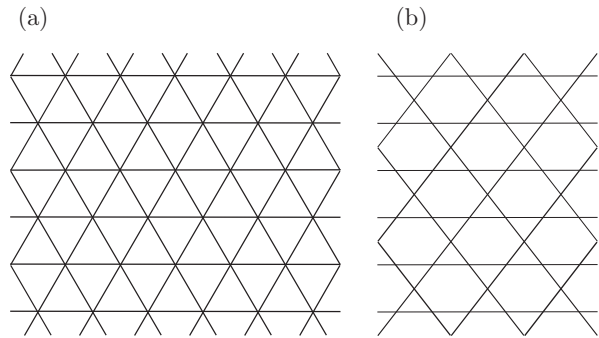}
\caption{
The illustration of the (a) triangular and the (b) kagome 
lattice.
}
\label{fig:fig2}
\end{center}
\end{figure}

The paper is organized as follows: 
Section II describes the model and the method. 
The results are presented and discussed in Section III. 
Section IV is devoted to the summary and discussions.  

\section{Model and Simulation Method}

We investigated the AFM Ising model 
with nearest-neighbor interaction on the pyrochlore lattice. 
Later, we also treat the Ising model on the triangular 
and kagome lattices. 
The pyrochlore lattice is illustrated in Fig.~\ref{fig:fig1}, 
whereas the triangular and kagome lattices are illustrated 
in Fig.~\ref{fig:fig2}.

The Hamiltonian is given by
\begin{equation}
  H = J \sum_{\l ij \r} s_i s_j, \quad (s_i = \pm 1), 
\end{equation}
where $\l ij \r$ stands for the nearest-neighbor pairs. 
Hereafter, the coupling $J$ is set as 1 unless otherwise specified.
We are especially interested in the site dilution of spins.
Then, the Hamiltonian becomes
\begin{equation}
  H = J \sum_{\l ij \r} c_i c_j s_i s_j, \quad (c_i = 1 \ {\rm or} \ 0). 
\end{equation}
Here $c_i$ is the quenched variable, and the concentration 
of vacancies is denoted by $x$. 

Quenched randomness is investigated basically under 
two different constraints, a grand-canonical constraint
(average density of vacancies fixed)
and a canonical constraint (total number of vacancies constant) 
\cite{Aharony,Marques}. 
Here, we use the canonical constraint for the site dilution. 
As for the dilution concentration $x$, we treat 
$x$= 0.0 (pure), 0.1, 0.2, 0.3, 0.4, 0.5, 0.6, 0.7, 0.8, 
and 0.9. 

To get precise numerical information on the entropy of the system,   
we use the WL method that directly calculates 
the energy DOS.
Let us briefly review the WL algorithm. A random walk
in energy space is performed with a probability proportional
to the reciprocal of the DOS, $1/g(E$), which results in a flat
histogram of the energy distribution. Actually, we move
based on the transition probability from energy level $E_1$ to $E_2$:
\begin{equation}
  p(E_1 \to E_2) = \min\big[1, \frac{g(E_1)}{g(E_2)}\big].
\end{equation}
Since the exact form of $g(E)$ is not known {\it a priori}, we
determine $g(E)$ iteratively. Introducing the modification factor
$f_i$, $g(E)$ is modified by
\begin{equation}
  \ln g(E) \to \ln g(E) + \ln f_i
\end{equation}
every time the state is visited. At the same time the energy
histogram $h(E)$ is updated as
\begin{equation}
  h(E) \to h(E) + 1.
\end{equation}
The modification factor $f_i$ is gradually reduced to unity by
checking the ``flatness" of the energy histogram. The ``flatness" 
is checked such that the histogram for all possible $E$ is not less
than some value of the average histogram, e.g., 80\%. Then $f_i$
is modified as
\begin{equation}
  \ln f_{i+1} = \frac{1}{2} \ln f_i,
\end{equation}
and the histogram $h(E)$ is reset. As an initial value of $f_i$, 
we choose $f_0 = e$; as a final value, we choose $\ln f_i = 2^{-24}$, 
that is, $f_{24} \simeq 1.00000006$.

\section{Results}
\subsection{pyrochlore lattice}

\begin{figure}
\begin{center}
\includegraphics[width=8.0cm]{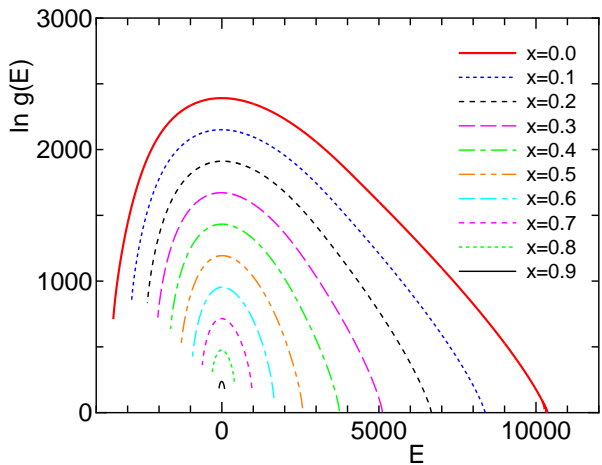}
\caption{
(Color online) 
The plot of $\ln g(E)$ as a function of $E$
of the Ising model on the pyrochlore lattice. 
The system size is $L=6 \ (N=3456)$. 
}
\label{fig:fig3}
\end{center}
\end{figure}

For the simulation of the Ising model on the pyrochlore lattice, 
we use the 16-site cubic unit cell of the pyrochlore lattice 
\cite{Shimaoka}, and the systems with $L \times L \times L$ 
unit cells with periodic boundary conditions are treated. 
We made simulations for the system sizes of $L$ =3, 4, 5, and 6; 
the numbers of sites are $N$ = 432, 1024, 2000, and 3456, 
respectively.

We first show the results of the Ising model on the pyrochlore lattice. 
In the WL algorithm, one directly calculates the ratio of $g(E)$ for different 
energies $E_1$ and $E_2$, $g(E_1)/g(E_2)$. If we are interested only in
the temperature dependence of the total energy or the specific heat, 
the ratio of $g(E)$ is enough. However, if we discuss the absolute value 
of entropy, the normalization of $g(E)$ is necessary.  In the case of the 
Ising model, each spin takes one of two states; thus, the normalization 
condition becomes
\begin{equation}
   \sum_E g(E) = 2^{N_{\rm spin}},
\end{equation}
where $N_{\rm spin}$ is the number of spins.  In the case of dilution, 
the number of spins $N_{\rm spin}$ is different 
from the number of sites $N$. 

We plot $\ln g(E)$, essentially the entropy, as a function of $E$ 
(in units of $J$) of the Ising model on the pyrochlore lattice, 
shown in Fig.~\ref{fig:fig3}. 
The system size is $L=6 \ (N=3456)$, 
and this is the plot of one sample for each $x$. 
The energy takes a value from $-NJ$ to $3NJ$ 
for the pure system ($x=0$). 
We note that the energy takes a value of the multiple of 
$4J$ for the pure system.  For diluted systems, 
the energy takes a value of the multiple of $J$, but 
the energy difference of a single spin flip is 
a multiple of $2J$. That is, the total energy is 
either the even number of $J$ or the odd number of $J$.

\begin{figure}
\begin{center}
\includegraphics[width=8.0cm]{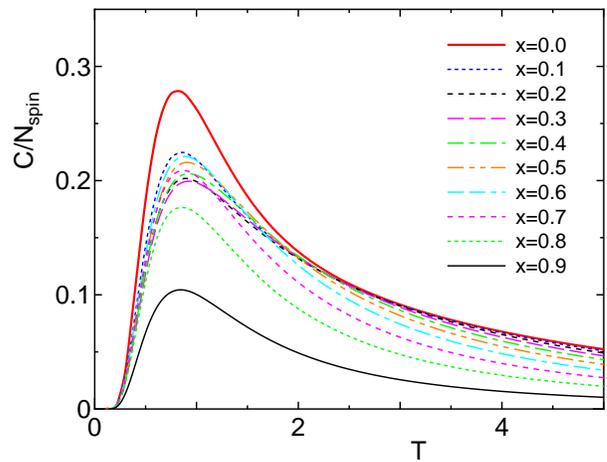}
\caption{
(Color online) 
The temperature dependence of the specific heat per spin 
of the AFM Ising model on the pyrochlore lattice. 
The system size is $L=6 \ (N=3456)$. 
}
\label{fig:fig4}
\end{center}
\end{figure}

The thermal average of a physical quantity $A$ 
at the inverse temperature, $\beta = 1/T$, is 
calculated from the knowledge of the energy DOS as
\begin{equation}
  \l A \r_{\beta} 
      = \frac{\sum_E e^{-\beta E} A(E) g(E)}{\sum_E e^{-\beta E} g(E)}.
\end{equation}
Then, the specific heat is calculated through the relation
\begin{equation}
  C = \frac{d \l E \r_{\beta}}{dT} 
   = \beta^2 (\l E^2 \r_{\beta}-\l E \r_{\beta}^2).
\end{equation}
The temperature dependence of the specific heat per spin 
of the AFM Ising model on the pyrochlore lattice 
($x = 0.0, 0.1, \cdots, 0.9$) is plotted in Fig.~\ref{fig:fig4}.  
The temperature $T$ is measured in units of $J$.
The system size is $L=6 \ (N=3456)$. 
The average was taken over 40 random samples. 
The statistical errors are smaller than the thickness of the curves.
We only show the data of $L=6$. For large enough system sizes of 
$L$ =4, 5, and 6, the size dependence is small, as in the scale of the 
plot in Fig.~\ref{fig:fig4}, 
because there is no phase transition associated with the long-range order.  
We should note that the peak of the specific heat becomes lower 
when $x$ is raised from 0.  

\begin{figure}
\begin{center}
\includegraphics[width=8.0cm]{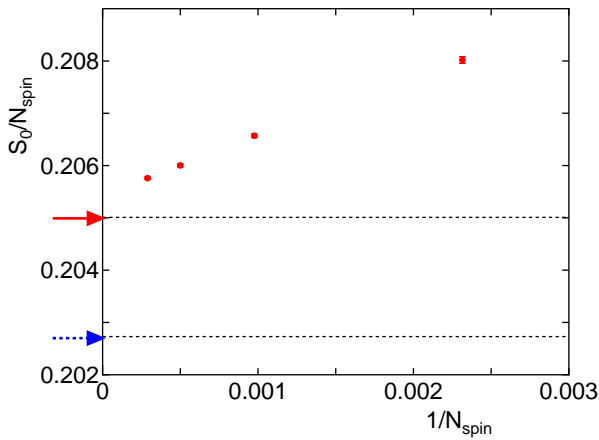}
\caption{
(Color online) 
The plot of the residual entropy per spin 
of the pure AFM Ising model on the pyrochlore lattice 
as a function of $1/N$  \ ($N=N_{\rm spin}$). 
The system size is $L=3 \ (N=432)$, 
$L=4 \ (N=1024)$, $L=5 \ (N=2000)$ and $L=6 \ (N=3456)$. 
The precise estimate by Nagle \cite{Nagle} (0.20501) is shown 
by the solid red arrow, whereas the approximation by Pauling 
\cite{Pauling} (0.20273) is shown by the dotted blue arrow.
}
\label{fig:fig5}
\end{center}
\end{figure}

Before discussing the thermal average of the entropy, 
we checked the accuracy of the calculation of the entropy 
in detail.  The residual entropy can be calculated from 
the raw data of $\ln g(E)$, which is shown in Fig.~\ref{fig:fig1}.  
We plot the residual entropy for the pure system ($x=0$) 
as a function of $1/N$ \ ($N=N_{\rm spin}$) in Fig.~\ref{fig:fig5}. 
The precise estimate of the residual entropy of a pure system ($x=0$) 
using the series method by Nagle \cite{Nagle} is 0.20501, 
which is close to Pauling's estimate 
$(1/2)\ln (3/2)=0.20273$ \cite{Pauling}.
We also observed that the residual entropy approaches 
the precise estimate by Nagle \cite{Nagle}. 
This is the same plot as Fig.~4 of Ferreyra 
{\it et al.} \cite{Ferreyra}.

\begin{figure}
\begin{center}
\includegraphics[width=8.0cm]{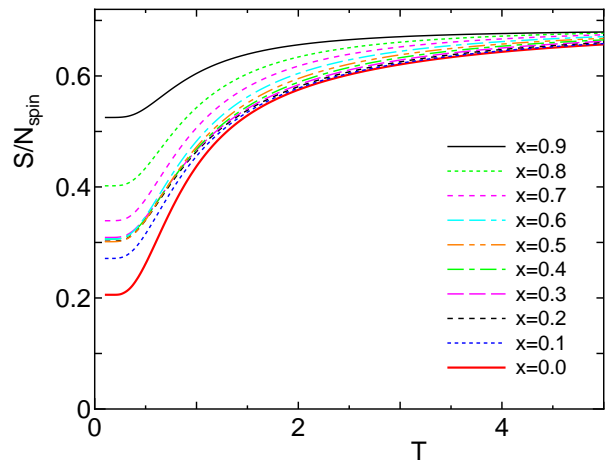}
\caption{
(Color online) 
The temperature dependence of the entropy per spin 
of the AFM Ising model on the pyrochlore lattice. 
The system size is $L=6 \ (N=3456)$. 
}
\label{fig:fig6}
\end{center}
\end{figure}

We calculate the thermal average of the entropy by
\begin{equation}
  S = \frac{\l E \r_{\beta}}{T} + \ln (\sum_E e^{-\beta E} g(E)).
\end{equation}
The temperature dependence of entropy per spin 
of the AFM Ising model on the pyrochlore lattice 
($x = 0.0, 0.1, \cdots, 0.9$) is plotted 
in Fig.~\ref{fig:fig6}. The average was taken over 40 samples. 
The system size is $L=6 \ (N=3456)$, and 
the size dependence is small as mentioned before.  
We note that, as $T \to 0$, the residual entropy becomes larger 
when $x$ is raised from 0. In the high-temperature limit 
where $T \to \infty$, the entropy per spin approaches $\ln 2=0.693$.

\begin{figure}
\begin{center}
\includegraphics[width=8.0cm]{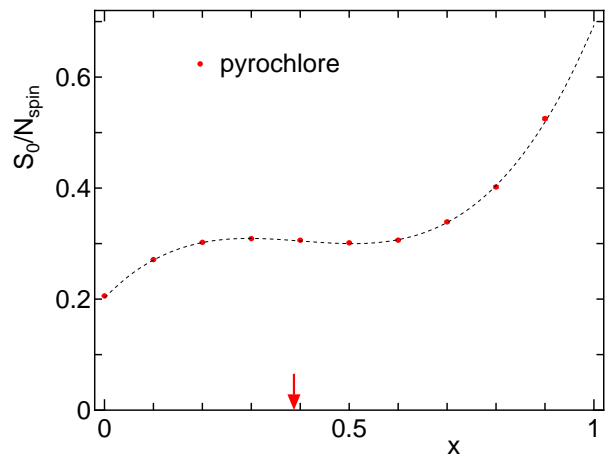}
\caption{
(Color online) 
The dilution concentration dependence of the residual entropy per spin 
of the AFM Ising model on the pyrochlore lattice. 
The system size is $L=6 \ (N=3456)$. 
The generalized Pauling approximation \cite{Ke} 
is also plotted by a dotted curve.
The site-percolation threshold of the pyrochlore lattice 
is shown by the red arrow.
}
\label{fig:fig7}
\end{center}
\end{figure}

\begin{table}
\caption{ 
The dilution concentration ($x$) dependence of residual entropy per spin 
for the pyrochlore lattice.
The system sizes are $L$ = 4, 5, and 6. 
The numbers in parentheses are the one-sigma uncertainty for the last digits.
}
\label{table1}
\begin{center}
\begin{tabular}{lllll}
\hline
\hline
$x$ \hspace*{0.4cm} & $L=4$ \hspace*{0.8cm} & $L=5$ \hspace*{0.8cm} 
& $L=6$ \hspace*{0.8cm} & generalized \\
\quad & ($N=1024$) & ($N=2000$) & ($N=3456$) & Pauling \cite{Ke}\\
\hline
0.0 & 0.2066(0)   & 0.2060(0)   & 0.2058(0)  &  0.2027 \\
0.1 & 0.2736(6)   & 0.2714(5)   & 0.2712(4)  &  0.2703 \\
0.2 & 0.3039(10)  & 0.3018(10)  & 0.3023(6)  &  0.3019 \\
0.3 & 0.3096(15)  & 0.3101(9)   & 0.3091(7)  &  0.3094 \\
0.4 & 0.3043(15)  & 0.3053(13)  & 0.3061(8)  &  0.3047 \\
0.5 & 0.2990(19)  & 0.3022(12)  & 0.3014(10) &  0.3000 \\
0.6 & 0.3084(24)  & 0.3065(14)  & 0.3063(10) &  0.3071 \\
0.7 & 0.3373(25)  & 0.3370(21)  & 0.3392(13) &  0.3380 \\
0.8 & 0.4048(30)  & 0.4025(25)  & 0.4021(19) &  0.4046 \\
0.9 & 0.5242(40)  & 0.5231(29)  & 0.5252(20) &  0.5190 \\
\hline
\end{tabular}
\end{center}
\end{table}

In Fig.~\ref{fig:fig7}, we plot the dilution concentration ($x$) 
dependence of the residual entropy per spin. 
The system size is $L=6 \ (N=3456)$. 
The average was taken over 40 random samples.
The statistical errors are smaller than the size of the marks. 
The size dependence is very small, and the size difference 
is not appreciable with this scale of plot. 
We observed the nonmonotonic concentration dependence of 
the residual entropy, which was experimentally 
reported \cite{Ke,Lin}. 
We also plot the generalized Pauling approximation \cite{Ke}:
\begin{eqnarray*}
   \frac{S_0(x)}{N_{\rm spin}} 
       &=& \ln 2 + 3\ln(1/2)(1-x)x^2 + 2\ln(3/4)(1-x)^2x \\ 
       &+& (1/2)\ln(3/8)(1-x)^3
\end{eqnarray*}
displayed by a dotted curve.
In order to show the numerical data explicitly,
we tabulate the measured values of the residual entropy 
for $L$=4, 5, and 6 in Table~\ref{table1}. 
The numbers in parentheses are the one-sigma uncertainty 
for the last digits, 
which was estimated by averaging over 40 samples.
The statistical errors for $x \ne 0$ come from 
the average over random samples, whereas they are only for 
random numbers of simulation for $x=0$, which are very small. 

The obtained values agree with the generalized Pauling 
approximation~\cite{Ke} to three digits for most of the range of $x$, 
indicating that it is essential for the argument based on the fraction 
of the configurations of a tetrahedron which satisfy 
the ground state condition. 
For the weak dilution region, the residual entropy becomes 
larger when $x$ is raised from 0. 
When one spin is missing from a tetrahedron, 
there are still 6 lowest-energy configurations 
out of the 8 possible configurations. 
This proportion is larger than the undeleted case 
that there are 6 lowest-energy configurations 
out of the 16 possible configurations 
in the tetrahedron. 
For the strong dilution limit ($x \to 1$), 
the residual entropy per spin approaches $\ln 2$
(=0.693), because all the spins become free. 
We conclude that 
the nonmonotonic dilution concentration dependence 
is primarily accounted for by the nearest-neighbor 
frustration interaction of the basic tetrahedron unit.

Analyzing the dilution concentration 
dependence of the specific-heat peak 
shown in Fig.~\ref{fig:fig4}, 
we also observed that the specific-heat peak 
shows the nonmonotic dependence. 
The peak decreases as a function of $x$ for small $x$, 
increases slightly, and then decreases again. 
It is anti-correlated with the behavior of the residual
entropy due to the increase of the entropy with the 
temperature, as calculated by
$$
\Delta S = \int \big( \frac{C}{T} \big) \ dT.
$$
This behavior of the specific-heat peak was 
previously reported by Ref.~\cite{Lin}.

\subsection{triangular and kagome lattices}

To elucidate the nonmonotonic dilution effects on the frustration 
of the pyrochlore lattice, a comparison with other frustrated 
systems will be interesting. 
For this purpose, we studied the AFM Ising model 
on the two-dimensional triangular lattice. 
The exact solution of this model was given by Wannier \cite{Wannier}, 
who established that this system has no long-range 
order due to frustration at all the temperatures. 
The residual entropy of the AFM Ising model 
on the triangular lattice was calculated to be 0.323066 \cite{Wannier}:
$$
  \frac{S_0}{N} = \frac{2}{\pi} \int_0^{\pi/3} \ln(2\cos \omega) 
                  \ d\omega = 0.323066.
$$

The dilution effects of the AFM Ising model 
on the triangular lattice was studied by Yao \cite{Yao} 
using the WL method.  However, a systematic 
study was not made for the dilution concentration dependence 
over the full range. 

We carried out the WL study of the AFM 
Ising model on the triangular lattice. 
We use the $L \times L$ system with the periodic conditions, 
and the system sizes are $L=24$ ($N=576$), $L=32$ ($N=1024$), 
and $L=48$ ($N=2304$). The simulation conditions 
were the same as those used on the pyrochlore lattice. 
We again averaged over 40 random samples for each size $L$ and 
for each dilution concentration $x$.

\begin{figure}
\begin{center}
\includegraphics[width=8.0cm]{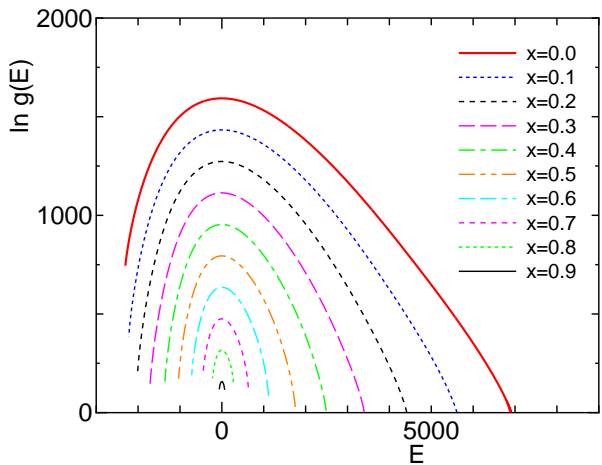}
\caption{
(Color online) 
The plot of $\ln g(E)$ as a function of $E$
of the Ising model on the triangular lattice. 
The system size is $L=48 \ (N=2304)$. 
}
\label{fig:fig8}
\end{center}
\end{figure}

We plot $\ln g(E)$ as a function of $E$ 
of the Ising model on the triangular lattice, 
shown in Fig.~\ref{fig:fig8}. 
The system size is $L=48 \ (N=2304)$, 
and this is the plot for one sample. 
The energy DOS $g(E)$ is 
normalized as $\sum_E g(E) = 2^{N_{\rm spin}}$, 
where $N_{\rm spin}$ is the number of spins. 
The energy takes a value from $-NJ$ to $3NJ$ 
for the pure system. 

\begin{figure}
\begin{center}
\includegraphics[width=8.0cm]{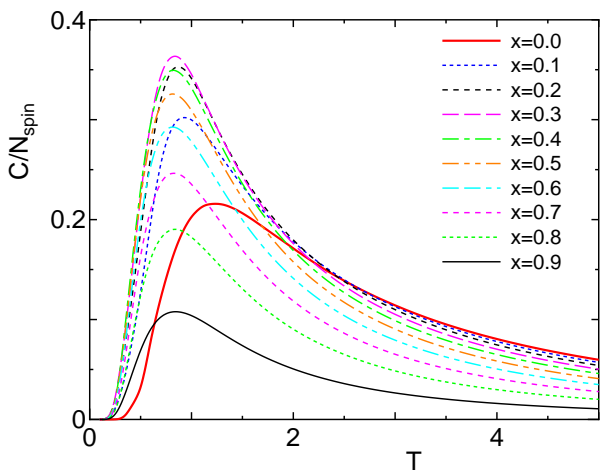}
\caption{
(Color online) 
The temperature dependence of the specific heat per spin 
of the AFM Ising model on the triangular lattice. 
The system size is $L=48 \ (N=2304)$. 
}
\label{fig:fig9}
\end{center}
\end{figure}

Using the data of the energy DOS, we plot the temperature 
dependence of the specific heat per spin 
of the AFM Ising model on the triangular lattice 
($x = 0.0, 0.1, \cdots, 0.9$), 
seen in Fig.~\ref{fig:fig9}.  We plot the data of the system size 
$L=48 \ (N=2304)$. The size dependence is very small for large 
enough sizes, $L=24$ ($N=576$), $L=32$ ($N=1024$), 
and $L=48$ ($N=2304$). 
The peak of the specific heat becomes higher when $x$ is raised from 0,
which was also shown by Yao \cite{Yao}.

\begin{figure}
\begin{center}
\includegraphics[width=8.0cm]{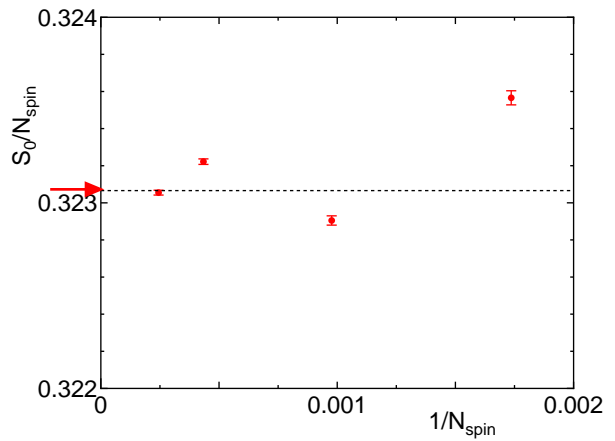}
\caption{
(Color online) 
The plot of the residual entropy per spin 
of the pure Ising model on the triangular lattice
as a function of $1/N$ \ ($N=N_{\rm spin}$). 
The system size is $L=24 \ (N=576)$, 
$L=32 \ (N=1024)$, $L=48 \ (N=2304)$, and $L=64 \ (N=4092)$. 
The exact value by Wannier \cite{Wannier} (0.323066) is shown 
by the red arrow.
}
\label{fig:fig10}
\end{center}
\end{figure}

The size dependence of the residual entropy of the pure 
AFM Ising model 
on the triangular lattice is given in Fig.~\ref{fig:fig10}. 
Wannier \cite{Wannier} exactly solved the AFM 
Ising model on the triangular lattice, and 
the residual entropy of a pure system ($x=0$) is 0.323066.
We plot $S_0/N_{\rm spin}$ as a function of $1/N_{\rm spin}$. 
Since there is an oscillatory behavior, we also plot 
the data for $L=64$, observing that the residual entropy per 
spin approaches Wannier's exact value in the limit as 
$N_{\rm spin} \to \infty$. 

\begin{figure}
\begin{center}
\includegraphics[width=8.0cm]{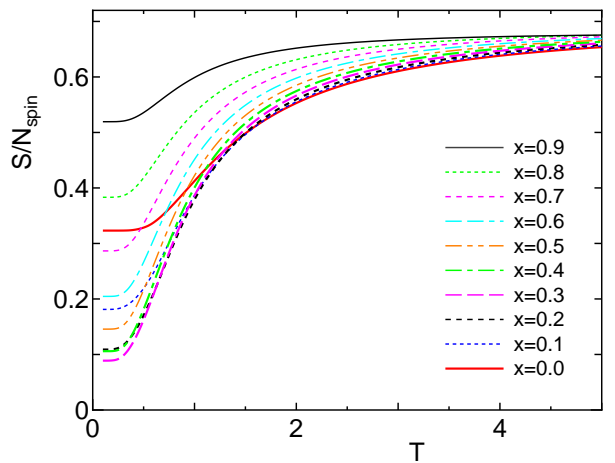}
\caption{
(Color online) 
The temperature dependence of the entropy per spin 
of the AFM Ising model on the triangular lattice. 
The system size is $L=48 \ (N=2304)$. 
}
\label{fig:fig11}
\end{center}
\end{figure}

The temperature dependence of the entropy 
of the AFM Ising model on the triangular lattice 
($x = 0.0, 0.1, \cdots, 0.9$) is plotted 
in Fig.~\ref{fig:fig11}. We plot the data of the system size 
$L=48 \ (N=2304)$. Here, the size dependence is very small and 
the residual entropy becomes smaller when $x$ is raised from 0, 
which was also shown by Yao \cite{Yao}.

\begin{figure}
\begin{center}
\includegraphics[width=8.0cm]{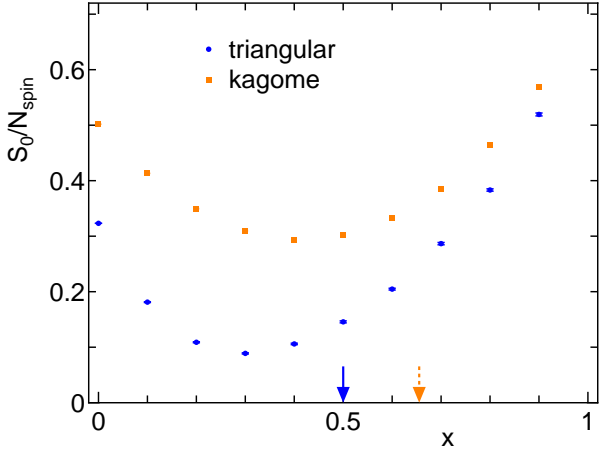}
\caption{
(Color online) 
The dilution concentration dependence of the residual entropy per spin 
of the AFM Ising model on the triangular lattice 
and the kagome lattice. 
The system size is $L=48$ ($N=2304$ for the triangular lattice 
and $N=3456$ for the kagome lattice). 
The site-percolation thresholds 
of the triangular and kagome lattices are shown by the 
solid blue 
and dotted orange arrows, respectively.
}
\label{fig:fig12}
\end{center}
\end{figure}

The dilution concentration ($x$) dependence 
of the residual entropy per spin of the AFM Ising model 
on the triangular lattice 
is given in Fig.~\ref{fig:fig12}.
The system size is $L=48 \ (N=2304)$. 
The residual entropy becomes smaller when $x$ is raised 
from 0, which is contrary to the results of the pyrochlore lattice. 
In the triangular lattice case, the frustration in 
the basic unit of the triangle is resolved 
if one spin is deleted from 3 sites. 
Since the frustration is partially resolved, 
the degeneracy of the ground states becomes smaller. 
In the case of the tetrahedron, on the contrary, there still remains 
a frustration even if one spin is deleted from 4 sites.
For the strong dilution limit 
($x \to 1$), the residual entropy per spin approaches $\ln 2$
(=0.693) because all the spins become free. 

To study another frustrated system, we also considered 
the AFM Ising model on the kagome lattice. 
The pyrochlore lattice can be viewed as an alternating 
stacking of kagome layers and sparse triangular layers. 
The macroscopic degeneracy of the kagome layers 
was studied when the magnetic field is applied 
along the [111] direction \cite{Matsuhira}.

The AFM Ising model on the kagome lattice 
was exactly solved by Kano and Naya \cite{Kano}. 
It was shown that this system has no long-range 
order due to frustration for all the temperatures. 
The residual entropy of the AFM Ising model 
on the kagome lattice was calculated to be 0.50183 \cite{Kano}:
\begin{eqnarray*}
  \frac{S_0}{N} &=& \frac{1}{24\pi^2} \int_0^{2\pi} \int_0^{2\pi} 
  \ln [21-4 (\cos \omega_1 + \cos \omega_2 + \\ &~& \cos(\omega_1+\omega_2))] 
  \ d\omega_1 d\omega_2 = 0.50183.
\end{eqnarray*}

In the WL study of the AFM 
Ising model on the kagome lattice, 
we treated the kagome lattice of $L \times (3/2)L$, and 
the system sizes are $L=24$ ($N=864$), $L=32$ ($N=1536$), 
and $L=48$ ($N=3456$).
We averaged over 40 random samples for each size $L$ and for each dilute 
concentration $x$.

\begin{figure}
\begin{center}
\includegraphics[width=8.0cm]{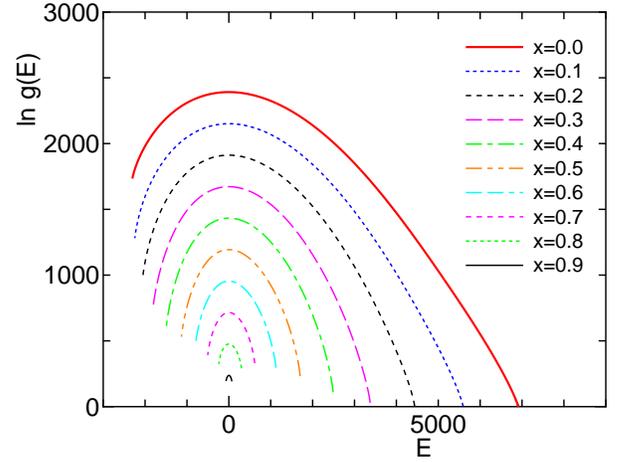}
\caption{
(Color online) 
The plot of $\ln g(E)$ as a function of $E$
of the Ising model on the kagome lattice. 
The system size is $L=48 \ (N=3456)$. 
}
\label{fig:fig13}
\end{center}
\end{figure}

We plot $\ln g(E)$ as a function of $E$ 
of the AFM Ising model on the kagome lattice, 
shown in Fig.~\ref{fig:fig13}. 
The system size is $L=48 \ (N=3456)$, 
and this is the plot for one sample. 
The energy DOS $g(E)$ is 
normalized as $\sum_E g(E) = 2^{N_{\rm spin}}$, 
where $N_{\rm spin}$ is the number of spins. 
The energy takes a value from $-(2/3)NJ$ to $2NJ$ 
for the pure system. 

\begin{figure}
\begin{center}
\includegraphics[width=8.0cm]{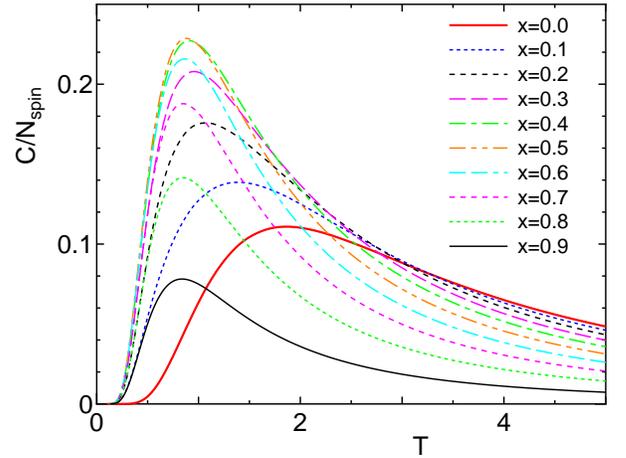}
\caption{
(Color online) 
The temperature dependence of the specific heat per spin 
of the AFM Ising model on the kagome lattice. 
The system size is $L=48 \ (N=3456)$. 
}
\label{fig:fig14}
\end{center}
\end{figure}

The temperature dependence of the specific heat 
of the AFM Ising model on the kagome lattice 
($x = 0.0, 0.1, \cdots, 0.9$) is plotted 
in Fig.~\ref{fig:fig14}. 
Averaging was performed over 40 samples. 
We plot the data for the system size 
$L=48 \ (N=3456)$. The size dependence is very small, 
and the peak of the specific heat becomes higher when $x$ is raised, 
same as in the triangular lattice. 

\begin{figure}
\begin{center}
\includegraphics[width=8.0cm]{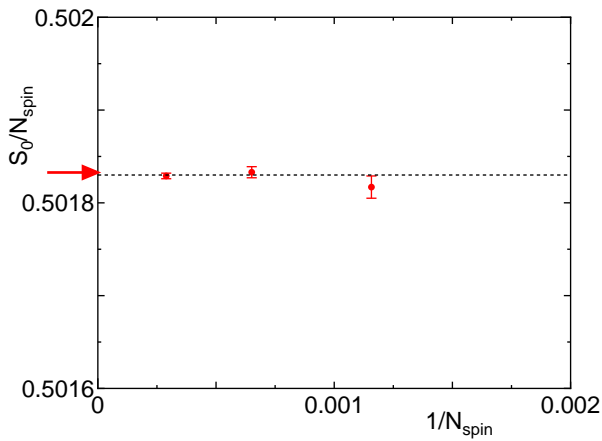}
\caption{
(Color online) 
The plot of the residual entropy per spin 
of the pure Ising model on the kagome lattice 
as a function of $1/N$ \ ($N=N_{\rm spin}$). 
The system size is $L=24 \ (N=864)$, 
$L=32 \ (N=1536)$, and $L=48 \ (N=3456)$. 
The exact value by Kano and Naya \cite{Kano} (0.50183) is shown 
by the red arrow.
}
\label{fig:fig15}
\end{center}
\end{figure}

To confirm the accuracy of our calculation, we plot the size dependence 
of the residual entropy as a function of $1/N_{\rm spin}$ 
for the pure AFM Ising model on the kagome lattice in Fig.~\ref{fig:fig15}. 
Our calculated result for the residual entropy of the AFM Ising model 
on the kagome lattice approaches the exactly calculated value 
given by Kano and Naya as 0.50183 \cite{Kano}.  

\begin{figure}
\begin{center}
\includegraphics[width=8.0cm]{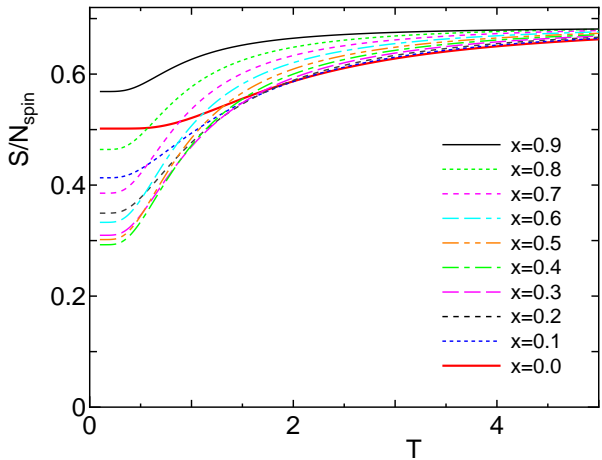}
\caption{
(Color online) 
The temperature dependence of the entropy per spin 
of the AFM Ising model on the kagome lattice. 
The system size is $L=48 \ (N=3456)$. 
}
\label{fig:fig16}
\end{center}
\end{figure}

The temperature dependence of the entropy 
of the AFM Ising model on the kagome lattice 
($x = 0.0, 0.1, \cdots, 0.9$) is shown 
in Fig.~\ref{fig:fig16}. We plot the data for the system size 
$L=48 \ (N=3456)$. The size dependence is very small, 
and the residual entropy becomes smaller when $x$ is raised from 0, 
exhibiting the same behavior as in the triangular lattice.

We plot the dilution concentration ($x$) dependence 
of the residual entropy per spin of the AFM Ising model 
on the kagome lattice, 
also shown in Fig.~\ref{fig:fig12}.
The system size is $L=48 \ (N=3456)$. 
The residual entropy becomes smaller when $x$ is raised 
from 0, the same as that of the triangular 
lattice. The frustration comes from a triangle in the case 
of the kagome lattice.  Thus, the frustration is resolved 
when one spin is deleted from this triangle. 

\section{Summary and Discussions}

We studied the diluted antiferromagnetic Ising model 
on the pyrochlore lattice.  
Using the Wang-Landau algorithm, which directly calculates 
the energy density of states, we calculated 
the entropy of the system with a high level of accuracy. 
We discussed the nonmonotonic dilution concentration dependence 
of the residual entropy, and obtained a very good comparison 
with the generalized Pauling approximation proposed 
by Ke {\it et al.} \cite{Ke}.

We also investigated other frustrated systems, 
the antiferromagnetic Ising model on the triangular lattice 
and the kagome lattice. 
We showed the difference in the dilution effects 
between the system on the pyrochlore lattice 
and that on other frustrated lattices. 
The dilution concentration dependence of the residual entropy 
was compared in Fig.~\ref{fig:fig7} and Fig.~\ref{fig:fig12}.
For the pyrochlore lattice, the residual entropy increases 
with $x$ for a weak dilution.  When one spin is deleted 
from the tetrahedron, there still remains the frustration 
from the other three spins.  However, for the triangular 
and kagome lattices, the residual entropy decreases with $x$. 
When one spin is deleted from the triangle, the frustration 
is resolved. 
The decrease of the residual entropy for the AFM Ising model 
on the triangular lattice up to $x \le 0.15$ was 
shown in Ref.~\cite{Yao}, 
but to the best of our knowledge, no reports have been made for the diluted 
AFM Ising model on the kagome lattice.

In Fig.~\ref{fig:fig7} and Fig.~\ref{fig:fig12}, 
we also give the values of the site-percolation thresholds.
The site-percolation thresholds of the pyrochlore lattice, 
the triangular lattice, and the kagome lattice are 
$0.39\cdots$ \cite{Henley}, 0.5 (=1/2), and $0.6527\cdots$ 
(= $1 - 2 \sin(\pi/18)$) \cite{Sykes}, respectively.
We may say that above the percolation threshold, 
the residual entropy increases and approaches 
the free-spin value, $\ln 2 = 0.691$.  
In the argument of the generalized Pauling 
approximation \cite{Ke}, only the local structure 
within the tetrahedron was considered, but it yields 
a good comparison with the numerical data. Thus, the percolation 
threshold is seen not to be directly related to the dilution 
concentration dependence of the residual entropy.

Based on the high-accurate calculation of the entropy, 
together with the comparative study of other 
frustrated systems, we conclude that 
the nonmonotonic dilution concentration dependence 
of the residual entropy of the pyrochlore lattice 
is primarily accounted for by the nearest-neighbor 
frustration interaction of basic tetrahedron unit.

The long-range dipole interaction may play a role 
in the detailed comparisons with experimental results. 
Lin {\it et al.}~\cite{Lin} reported the comparative 
studies of the results of experiments and (canonical) 
Monte Carlo simulations, which showed a difference 
between Dy and Ho systems. 
There was still a discrepancy between the experimental 
and Monte Carlo results.
An accurate Wang-Landau study on 
the pyrochlore system that includes the next nearest-neighbor 
interaction is still needed, and it will be left to a future work. 

Quite recently, the topological spin glass behavior 
has been discussed for diluted spin ice \cite{Sen}. 
The relevance of our results to their study will be interesting, 
but for now is an open question.

\section*{Acknowledgment}

We thank Vitalii Kapitan and Hiromi Otsuka for valuable discussions. 
The computer cluster of Far Eastern Federal University 
was used for computation. 
This work was supported by a Grant-in-Aid for Scientific Research 
from the Japan Society for the Promotion of Science, 
Grant Nos. JP25400406, JP16K05480. 
This work was also financially supported by a grant of the President 
of the Russian Federation for young scientists and graduate students, 
in accordance the Program of development priority direction 
``Strategic information technology, including the creation 
of supercomputers and software development".

\end{document}